# ON POSSIBLE WAYS FOR FUTURE REDEFINITION OF THE SI TIME UNIT

Viacheslav V. Khruschov

*Russian Research Institute of Metrological Service,, Moscow, Russia, e-mail: khkon@vniims.ru*

Abstract. The current methods for realization of the second with the highest accuracy through the frequency of the hyperfine transition between two levels of the ground state of the cesium 133 atom are considered. It is discussed possible ways for the future redefinition of the SI time unit with the help of optical frequency standards, as well as thorium and super-high frequency standards. Several constants which can be also used for definitions of the time, mass and electric current units are analyzed.

*Keywords:* second**,** redefinition of the SI units; defining constant; cesium frequency standard; optical frequency standard; thorium frequency standard, permeability of free space; atomic mass unit.

At present the International System of Units (SI) is used for measurements for research and needs of industry, technologies and trade. The New SI was approved at the XXVI CGPM in November 2018 and started to put into practice beginning with 20 May, 2019. In fact only four base units (the kilogram, the ampere, the mole and the kelvin) were redefined by way of fixed values of some physical constants (PC) as it was done for redefinition of the ampere in 1946 and the metre in 1983 [1-3]. The new term was introduced in the 9[th] edition of the SI brochure [4], namely, "the defining constant" (DC) for the usual term "PC with fixed value". The accepted New SI will undoubtedly improve the quality of measurements and the guarantee of their unity.

It is desirable to keep on improving of the SI taking into account the rapid development of science and modern technologies. The elaboration of a new definition of the time unit (the second) is one of the main line of SI improving. It is known that the fixed value of the frequency of the superfine transition in the ground state of the cesium 133 atom is used for the up-to-date second definition. This definition was adopted at the XIII CGPM in 1967. The wording of this definition was changed for years (including the XXVI CGPM in 2018). Now the following definition of the second is used [5].

*The second, symbol s, is the SI unit of time. It is defined by taking the fixed numerical value of the cesium frequency $\Delta\nu_{Cs}$, the unperturbed ground-state hyperfine transition frequency of the cesium 133 atom, to be 9 192 631 770 when expressed in the unit Hz, which is equal to $s^{-1}$.*

The SI is thus defined by fixing fundamental constants, among which $\Delta\nu_{Cs}$. *The International System of Units, the SI, is the system of units in which the unperturbed ground state hyperfine transition frequency of the cesium 133 atom $\Delta\nu_{Cs}$ is 9 192 631 770 Hz.*

Ever-developing technology progress made possible enhancing the accuracy of the conservation and transmission of the second with the help of the current definition. Now the limit of accuracy is achieved and is of the order of $10^{-16}$ (usually the indicator of accuracy is the relative non excluded systematic uncertainty or its inverse value). It follows that a redefinition of the time unit is worth consideration. Its adoption appears to be at the XXVIII CGPM in 2026. In 2016 the CCTF accepted the CCTF-roadmap with the description of the strategy for transition to a new definition of the second basing on optical frequency standards [6]. The optical frequency standards possess very high accuracy and stability at reproducing the transition frequency value for a selected atom as well as convenience at signal transmissions with the optical fiber techniques. One may consult CCTF publications about problems associated with the realization of the strategy for transition to a new definition of the second basing on optical frequency standards [7].

The present paper is concerned with some peculiarities of optical frequency standards as well as thorium and ultra-high-frequency standards for comparison estimates of their likely uses in coming and future procedures of a redefinition of the second. Sets of DCs are also considered, which can be used in redefinition of some SI units including the unit of time.

A short outline of the paper is as follows. First we consider the current methods for realization of the definition of the second on the base of the unperturbed ground-state hyperfine transition frequency of the cesium 133 atom. Then we discuss some peculiarities for a redefinition of the second on the base of optical frequency standards, thorium and ultra-high-frequency standards. It is also suggested a DC set for likely future definitions of the SI units including the second, kilogram and ampere.

**Current methods for realization of the definition of the second with the highest accuracy.** The redefinition of the kilogram, ampere, mol and kelvin was approved at the XXVI CGPM in 2018 with the help of the fixed values of the Planck constant ($h$), the electron charge magnitude ($e$), the Avogadro constant ($N_A$) and the Boltzmann constant ($k$), correspondingly [5]. The definition of the second holds in fact, but it takes another wording on the base of the DC value (see above). However any realization of this definition always happens with an uncertainty that depends on a used device. Nowadays atomic frequency standards (AFS) are used for this purpose. The most-used AFS are the standards on the base of unperturbed ground-state hyperfine transition frequencies of the cesium, rubidium and hydrogen atoms [8].

Atoms are cooled with laser radiation in AFS, for instance, their thermal velocities in cesium and rubidium fountains are of the order of a few cm/s [8]. Thereby systematic uncertainties through the Doppler effect and thermal radiation are reduced. A width of the principal line of radiation is of the order of 1 Hz. Hydrogen masers offer their high stability, which is characterized with the Allan index about $10^{-15}$ or less. The Allan index depends on relative statistical uncertainties, it decreases when a measurement time increases.

The primary reproduction of the second is now carried out with the help of the AFS on the base of cesium fountains with the accuracy and instability of the order of $10^{-16}$ [8]. In addition, it is attained the high accuracy of comparison of frequencies between cesium and rubidium fountains, so the rubidium fountains can be used for the reproduction of the second as well. Nevertheless, high metrological characteristics of cesium and rubidium fountains, which are based on the cesium standard, reached their limits. Thereby, transition to new frequency standards is needed for further achievements in time measurements. Optical frequency standards (OFS) are main candidates for this purpose.

**Methods for redefinition of the SI time unit on the base of optical and thorium frequency standards.** Recently the important progress is occurred in creation of lasers with high accuracy and stability, which are many times (by a factor of the order of $10^2$ or higher) superior to accuracy and stability of lasers based on the cesium standard. Very narrow resonances have been obtained due to increasing of interaction times with separate electromagnetic fields, two-photon absorption without Doppler widening for atoms and ions captured into traps. This permits to lower a level of relative systematic uncertainty at reproduction of a laser frequency to $10^{-18}$. Lasers of this type are main constituents of OFS [9, 10].

A component of the spectrum of an OFS laser must coincide with a known spectral line of an atomic or molecular gas; therefore it must be as narrow as possible to attain high accuracy and stability. Forms of atomic and molecular transition lines are found with laser combs. A periodical sequence of impulses generated with the laser contains a set of equidistant frequencies that makes it possible to measure frequency values in a radio-frequency to ultraviolet range. Later on OFS development will be depend on working out of methods of generation of very narrow and stabile resonances with the help of single ions cooled with laser radiation and confined into electromagnetic traps such as Penning traps or radio-frequency quadrupole traps. For instance, one of the best frequency standards consist of a mercury ion captured in a radio-frequency trap with an internal size about 1 mm and a laser comb. Attained in particular due to a very simple working scheme the relative systematic uncertainty of frequency value is of the order of $10^{-18}$ [9, 10].

Diminishing of instability at frequency reproduction to the level of the order of $10^{-16}$ is one of the main OFS qualities, moreover instability decreases when measurement time increases. In order one can optimally obtain high characteristics of the OFS non excluded systematic uncertainties must be less or of the order of statistical uncertainties. OFS using reveals unique opportunity for creation of a new standard of the time-frequency unit and its application in science and technologies, for instance, for high-accuracy PC measurements and search of their variations [11]. The best accuracy of atomic OFS is now achieved on the base of as neutral atoms as ions such as ytterbium-171, strontium-87 and mercury-199 atoms, as well as aluminium-27, ytterbium-171, and mercury-199 ions [8-10].

It should be said about possible frequency standards with optical frequencies of nucleus transitions [12]. At present it is known about possible optical transitions in thorium, uranium and protactinium nuclei. In so far as laser technologies with high intensity and narrow resonance widths of radiation can be developed only for photons with energies less than 100 eV, it is needed to investigate transitions between the ground and isomer states in thorium-229 [13, 14], uranium-235 [15] and protactinium-229 [16] nuclei. The most suitable frequency standard is the thorium-229 standard. The difference between energies of the ground and isomer states of the thorium nucleus is of the order of 8 eV. The relative non excluded systematic uncertainties about $10^{-19}$ can be obtained in the thorium standards [12, 14]. Single thorium ions confined in the Paul trap or thorium ions implemented in a crystal lattice are usually used. A radiative transition has not been detected yet between the ground and isomer states of the thorium nucleus, only a nonradiative transition has been detected with electrons of internal conversion [12].

**A set of constants for future definitions of the time, mass and electric current units.** Numerous discussions had been happened about advantages and disadvantages of various versions of new definitions of the SI units on the base of DC values in the lead-up to the XXVI CGPM in November 2018 (see e.g. [2, 3, 17-21]). The most preferable variant was the set consisting of $h$, $e$, $k$ and $N_A$ [2], this set had been adopted. As mentioned above the definition of the second on the base of the unperturbed ground-state hyperfine transition frequency of the cesium 133 atom (that is the DC value) retains its validity in fact. However at present the problem of a redefinition of the second is set for reason of reaching the limits in accuracy and stability for AFS based on the cesium standard. The versions of a redefinition of the second on the base of the OFS along with the thorium frequency standard are discussed in the previous section. A future redefinition of the second with the help of a fundamental constant with dimensions of mass ($\mu$) is considered in this section. As noted below it is possible to use as well a definition of the kilogram on the base of $\mu$ and a definition of the ampere on the base of the fixed value of the permeability of free space $\mu_0$.

It was considered previously in the works [17, 22] the definitions of the kilogram, mole, ampere and kelvin using the fixed values of $h$, $N_A$, $k$ and $\mu_0$ (the so-called system D, see below). Advantages and disadvantages of this system had been noted as compared with order systems among which are the system A (in use until 2019). The definitions of the kilogram, mole, ampere and kelvin are based on the fixed values of the International prototype of the kilogram ($m(K)$), $\mu_0$, the triple point of water ($T_{tpw}$) and the molar mass of carbon $^{12}$C ($M(^{12}C)$).

It should be noted all considered versions of the New SI are based on fixed values of the constants belonged to the set $h$, $e$, $k$, $N_A$, $m_u$, $\mu_0$, where $m_u$ –is the atomic mass unit. Taking into account that the values $k$ and $N_A$ are present in all versions one can classify an each version with picking out only two DC from $h$, $e$, $m_u$ and $\mu_0$. For example, the system B contains $h$, $e$; the system C contains $m_u$, $e$; the system D contains $h$, $\mu_0$; the system E contains $m_u$, $\mu_0$. The pair $h$, $m_u$ had been excluded in the work [17] since it was related to a redefinition of the second, as the pair $e$, $\mu_0$ had been excluded since it brought no advantages as compared with the versions B, C, D and E. But the pair $h$, $m_u$ does is of interest in this work.

Let us consider the pair $h$, $m_u$ for a redefinition of the second. The ratio of $h$ to $m_u$ divided by $c^2$ gives the quantity with dimensions of time which can be accepted for the unit of time. One may decide on the electron mass $m_e$ or any PC with dimensions of mass instead of $m_u$. Then a selected quantity should be named as the fundamental mass $\mu$. However at present this definition of the time unit cannot be realized despite of its clarity and simplicity. Really, the electromagnetic radiation used in the cesium standard is of the order of $10^{10}$ Hz, in the OFS is of the order of $10^{15}$ Hz, whereas laser combs with the frequency of the order of $10^{24}$ Hz or $10^{21}$ Hz are needed for realization of the new definition of the second, correspondingly. Creation of facilities of such type (super-high frequency standards) can be considered in the very distant outlook.

It is known the system B($h$, $e$, $k$, $N_A$) had been adopted at the XXVI CGPM in particularly due to the enhancement in accuracy of electromagnetic measurements. The zero uncertainty of the Josephson and von Klitzing constants, $K_J$ and $R_K$, is given by definition in the system B. Now, these constants can be used as the SI standards, thus the problem of the practical electromagnetic metrology is resolved, when presence of the fixed $K_J$ and $R_K$ accepted in 1990 was outside the SI. The $K_J$ and $R_K$ values accepted in 1990 had been substituted in the New SI on the base of the recent $h$ and $e$ values from the CODATA data [23]. However, $\mu_0$ and $M_u$ (the molar mass constant) must be changed according with the latest experimental data. For instance, this dependence for $\mu_0$ can be specified by $4\pi \times (1+\delta) \times 10^{-7}$ N A$^{-2}$, where $\delta = (\alpha/\alpha_* - 1)$, $\alpha_*$ is the latest experimental $\alpha$ value at the moment of redefinition ( a relative uncertainty $u_r$ for $\delta$ identical with $u_r$ for $\alpha$) [17]. Certainly, it is no problem concerning the system B at practical measurements in widespread use when deviations of $\mu_0$ and $M_u$ from $4\pi \times 10^{-7}$ N A$^{-2}$ and 1g mol$^{-1}$ are considerably small.

However in this work the system D based on $h$, $\mu_0$, $k$, and $N_A$ is chosen as preferable [22]. The system D had been proposed for the first time by the Working Group on the base SI units and fundamental constants of the Paris Academy of Science (in particular, it was proposed to fix the Planck charge $q_P = (2\varepsilon_0 hc)^{1/2}$) [24]. In the system D the definitions of the mole and kelvin are based on the fixed $N_A$ and $k$ values, the kilogram – on the fixed $h$ value, the ampere – on the fixed $\mu_0$ value. But the experimental uncertainties for the $K_J$ and $R_K$ values are occurred in the system D due to a varying $e$ value. These uncertainties are very small (~$10^{-10}$) so they can be neglected at practical measurements in widespread use. However, for high-precision measurements they should be included in systematic uncertainties. The fine-structure constant $\alpha$ value is of first importance for determination of $K_J$ and $R_K$ uncertainties. For example, the $\alpha$ value had been recently found without resorting to the quantum electrodynamics on the base of result of the high-precision experiment through atomic recoil ($h/m_u$ ($^{87}$Rb) measurement [25]). The measurements of such type will recur in the future so it will lead to increasing of accuracy of the $K_J$ and $R_K$ values. The $K_J$ and $R_K$ values accepted at the XXVI CGPM can be taken as the primary values.

Using the fixed value of $\mu_0 = 4\pi \times 10^{-7}$ N A$^{-2}$ is an apparent advantage of the system D. The $\mu_0$ fixation was realized at the redefinition of the ampere in 1946, and then the $c$ fixation was approved at the redefinition of the metre in 1983. The permittivity and permeability of free space, $\varepsilon_0$ and $\mu_0$, obey the relation $\varepsilon_0 \mu_0 c^2 = 1$ thus $\varepsilon_0$ has the fixed value $\varepsilon_0 = 8.854187817 \ldots \times 10^{-12}$ F m$^{-1}$ as for the system A. Then $u_r(e^2) = u_r(\alpha)$, taking into account $\alpha = e^2/(2\varepsilon_0 h c)$. Note that results of spectroscopic measurements depend on the units of length and time; it is of practical importance for evaluation of uncertainties of constants in the system D. For instance, one can ex-press $m_u$ and $e$ through the fixed constants $h$, $\mu_0$ and the metered constants $\alpha$, $A_r(e)$ and $R_\infty$. The fixed value of $\mu_0$ causes fixing of the vacuum impedance $Z_0$. So it is possible to find $R_K$ by comparison with $Z_0$ using the rated condenser, as well as $K_J$ with the help of the watt-balance without verification of their relations with $e$, $h$ and $\alpha$ [22]. Besides, the ampere is defined by the same procedure as in the system A.

The fixation of $h$ and $\mu_0$ accounts for the vacuum stability with respect to quantum phenomena and is an added reason for choosing the system D($h$, $\mu_0$, $k$, $N_A$) [22]. In this system α variation along with a variation of the interaction distance is accounted for by a variation of the electron charge magnitude $e$ ($\alpha^{-1} \sim 137$ at low energy and $\alpha^{-1} \sim 128$ on the scale of the Z boson mass [23]). So the system D is of considerable metrological importance.

The system D($h$, $\mu_0$, $k$, $N_A$) can be complemented with the fundamental mass $\mu$ or the system E with the fundamental mass $\mu$, i.e. E($\mu$, $\mu_0$ $k$, $N_A$), can be complemented with the Planck constant $h$; the system ($h$, $\mu$, $\mu_0$, $k$, $N_A$) is obtained in the end. If the speed of light in vacuum $c$ is added, an universal set of six DC ($c$, $h$, $\mu$, $\mu_0$, $k$, $N_A$) will be formed (the system F). The system F can be used for definition of six base units, namely, the metre, the second, the kilogram, the ampere, the kelvin, and the mole. In the future, the kilogram can be defined on the base of $\mu$. The adopted at the XVI CGPM in 1979 definition of the candela holds in fact.

**Conclusions.** The suitable variants of a new definition of the SI time unit, the second, have been considered in this paper. At present the most preferential variant is the definition of the second on the base of the OFS, moreover a family of optical standards can be used for a determination of their frequency-weighted average [26].

Different DC sets have been analyzed in detail for the purpose of presenting a relevant set for a redefinition of the second, which corresponds as possible to the DC sets of the SI and the New SI. The universal set of six DC ($c$, $h$, $\mu$, $\mu_0$, $k$, $N_A$) is well suited for this purpose. However, a realization of the new definitions based on the universal set can be performed in full extent only in the very distant outlook.